*Review*

# Cognitive Assessment and Training in Extended Reality: Multimodal Systems, Clinical Utility, and Current Challenges

Palmira Victoria González-Erena[1,2], Sara Fernández-Guinea[2], and Panagiotis Kourtesis[1-5]*

[1] Department of Psychology, The American College of Greece, 15342 Athens, Greece
[2] Department of Experimental Psychology, Cognitive Processes and Speech Therapy, Complutense University of Madrid, 28223 Madrid, Spain
[3] Department of Informatics & Telecommunications, National and Kapodistrian University of Athens, 16122, Athens, Greece
[4] Department of Psychology, National and Kapodistrian University of Athens, 157 84 Athens, Greece
[5] Department of Psychology, The University of Edinburgh, Edinburgh EH8 9Y, UK
* Correspondence: pkourtesis@acg.edu

**Abstract:** Extended Reality (XR) technologies—encompassing Virtual Reality (VR), Augmented Reality (AR), and Mixed Reality (MR)—are transforming cognitive assessment and training by offering immersive, interactive environments that simulate real-world tasks. XR enhances ecological validity while enabling real-time, multimodal data collection through tools such as galvanic skin response (GSR), electroencephalography (EEG), eye tracking (ET), hand tracking, and body tracking. This allows for a more comprehensive understanding of cognitive and emotional processes, as well as adaptive, personalized interventions for users. Despite these advancements, current XR applications often underutilize the full potential of multimodal integration, relying primarily on visual and auditory inputs. Challenges such as cybersickness, usability concerns, and accessibility barriers further limit the widespread adoption of XR tools in cognitive science and clinical practice. This review examines XR-based cognitive assessment and training, focusing on its advantages over traditional methods, including ecological validity, engagement, and adaptability. It also explores unresolved challenges such as system usability, cost, and the need for multimodal feedback integration. The review concludes by identifying opportunities for optimizing XR tools to improve cognitive evaluation and rehabilitation outcomes, particularly for diverse populations, including older adults and individuals with cognitive impairments.

## 1. Introduction

Extended Reality (XR), encompassing Virtual Reality (VR), Augmented Reality (AR), and Mixed Reality (MR), has transformed cognitive assessment and training by offering immersive, dynamic environments that simulate real-world tasks [101]. Traditional neuropsychological tests—such as paper-and-pencil tasks or static computerized exercises—often isolate cognitive functions under artificial conditions [44]. XR, in contrast, integrates real-world complexity into cognitive assessments and training [100]. For example, while a traditional memory test might involve recalling a list of words, XR can simulate a virtual shopping mall where participants must locate items on a list, recall their positions, and manage realistic distractions [63]. This not only evaluates memory but also incorporates attention, spatial navigation, and decision-making, offering a more ecologically valid reflection of real-world cognitive performance [63, 95].

A key innovation of XR is its ability to combine immersive, interactive experiences with multimodal feedback systems such as eye tracking (ET), galvanic skin response (GSR), electroencephalography (EEG), and body tracking [4, 54]. These technologies enable the real-time collection of behavioral, physiological, and neural data, providing deeper insights into cognitive and emotional states during task performance [54, 86]. For instance, in an XR-based attention task, eye-tracking data can reveal visual attention patterns, while EEG signals can indicate changes in cognitive load or mental fatigue [4, 114]. This continuous, multimodal data collection represents a significant advancement over traditional methods, which often capture only static performance snapshots [101].

XR also addresses challenges such as disengagement and the learning effect observed in repetitive cognitive tasks [100]. Immersive XR environments enhance user engagement and motivation, particularly for older adults and individuals with cognitive impairments, where adherence to training programs is often a concern [104, 112]. Additionally, XR systems can adapt task difficulty dynamically in real time based on user performance and cognitive load, ensuring personalized assessments and training programs aligned with individual abilities [27, 114].

In clinical contexts, XR demonstrates utility in assessing and rehabilitating cognitive impairments associated with neurodegenerative diseases, brain injuries, and neurodevelopmental disorders [20, 111]. For example, XR-based neuropsychological batteries can simulate daily activities, such as navigating virtual cities or managing household responsibilities, providing clinicians with ecologically valid insights into cognitive performance [42, 43]. XR's ability to collect longitudinal, personalized data further enhances its role in monitoring progress and tailoring interventions for improved cognitive outcomes [126].

Despite its potential, XR technologies face several challenges, including the underutilization of sensory modalities beyond visual and auditory feedback, issues such as cybersickness, and barriers to accessibility and usability [64, 101]. Addressing these limitations is essential for XR to fulfill its promise as a transformative tool in cognitive science.

### 1.1. Aim and Approach of This Narrative Review

This narrative review aims to explore the applications, benefits, and challenges of XR technologies in cognitive assessment and training. Specifically, it examines XR's contributions to enhancing ecological validity, integrating ET, EEG, GSR, and body tracking, and supporting clinical interventions. The review also identifies unresolved challenges, including usability concerns, cybersickness, and regulatory barriers, while proposing future directions for research and implementation.

To ensure a comprehensive and focused analysis, the literature was identified and selected based on the following approach:

- Databases searched: PubMed, IEEE Xplore, and Scopus.
- Keywords: "Extended Reality," "Cognitive Assessment," "Adaptive Systems," "Neuropsychological Testing," "XR-based Cognitive Training," "Eye Tracking," "EEG," "GSR," and "Body Tracking."
- Inclusion criteria: Experimental studies, theoretical frameworks, and applications of XR technologies in cognitive science published within the last 10 years.
- Focus: The review prioritizes studies on XR-based tools for cognitive assessment and rehabilitation, emphasizing the integration of ET, EEG, GSR, and body tracking data streams across diverse populations, including healthy individuals, older adults, and clinical groups with neurological or neurodevelopmental conditions.

By synthesizing findings from these sources, this review highlights XR's comparative advantages over traditional cognitive tools, its role in creating ecologically valid and adaptive assessments, and its potential for advancing cognitive training and rehabilitation.

## 2. Ecological Validity in XR-based Cognitive Assessment and Training

Ecological validity refers to the degree to which cognitive assessments reflect real-world scenarios, ensuring that findings gathered in controlled environments generalize to everyday functioning [25]. Traditional assessments, such as paper-and-pencil or static computerized tasks, often isolate cognitive functions like memory, attention, and problem-solving but fail to replicate the complexity and unpredictability of real-world tasks [44]. Cognitive performance is context-dependent and influenced by environmental, social, and situational factors [95]. This is particularly important in neuropsychology and cognitive rehabilitation, where assessments guide interventions aimed at improving real-world outcomes [31].

XR, including VR and MR, enhances ecological validity by simulating dynamic, real-world-like conditions that mirror daily challenges [100]. Unlike traditional assessments, XR allows participants to engage in tasks that require the simultaneous use of multiple cognitive functions—such as attention, memory, executive function, and visuospatial reasoning—within realistic scenarios [28, 63]. For example, a participant navigating a virtual city must make decisions, solve problems, and interact with the environment [51, 63]. This holistic approach replicates cognitive demands encountered in everyday life, making XR a more relevant and functional tool for assessment [100].

*2.1. Examples of XR-based Cognitive Tasks with Real-World Relevance*

Several XR-based tasks have been developed to assess cognitive abilities in immersive, contextually relevant settings [10, 83]. These tasks include:

- Memory: XR environments evoke autobiographical memories and enhance episodic memory by embedding tasks within realistic contexts, such as navigating a virtual home or completing a shopping list [55, 108].
- Prospective Memory: XR tasks replicate real-world challenges, such as remembering to perform time- or event-based actions (e.g., managing a virtual household or navigating a virtual shopping trip), offering a more naturalistic approach than lab-based button-press tasks [13, 58, 60, 72].
- Executive Function: Tasks like planning routes through virtual environments or multitasking in a simulated workspace mirror real-world problem-solving and adaptability demands [51, 63].
- Language: XR assessments integrate virtual avatars and multimodal feedback (e.g., eye gaze, speech, and facial expressions) within interactive social scenarios, providing insights into natural language comprehension and production [45, 97].
- Attention and Spatial Cognition: XR simulates complex tasks such as driving, cooking, or office work, requiring users to sustain attention, shift focus, and navigate spatially challenging environments [84, 118].

By replicating real-world cognitive demands in a controlled environment, XR enhances both the ecological validity and accuracy of cognitive assessments, making them more reflective of daily-life challenges [63].

## 3. Usability, Acceptability, and User Experience in XR

*3.1. Usability of XR Devices and Systems for Cognitive Assessment*

Usability is a critical factor for the successful implementation of XR systems in cognitive assessment, as tasks require sustained engagement and active participation [100]. Usability encompasses the ease with which individuals interact with XR hardware, such as headsets and sensors, as well as software interfaces that enable immersive experiences [22]. Poor usability—manifesting as unintuitive interfaces or uncomfortable hardware—can increase cognitive load and negatively impact task performance, compromising the accuracy and validity of assessment [2, 47, 69].

Significant advancements in XR usability have been achieved in recent years. Modern headsets are lighter, more ergonomic, and feature improved interactions through hand tracking, voice commands, and eye-tracking technologies [101, 120]. These advancements reduce the need for users to learn complex controls, allowing them to focus on tasks rather than managing system intricacies [73, 115]. Nevertheless, challenges remain, particularly for older adults and inexperienced users, who may find XR systems intimidating or difficult to navigate [48]. This intimidation can reduce their willingness to engage with XR-based assessments or training [22]. Additionally, individuals with cognitive impairments face unique barriers, as complex interfaces can pose significant challenges to their effective participation [93]. Addressing these usability concerns requires designing XR systems that prioritize simplicity, inclusivity, and accessibility, ensuring broader acceptance across diverse populations [101].

### 3.2. User Acceptability and Experience Across Various Populations

User acceptability refers to the extent to which XR systems are adopted and embraced across different populations, which is strongly influenced by user experience (UX)—the overall comfort, satisfaction, and engagement individuals derive from interacting with XR devices and applications [2]. A positive UX is crucial for the success of XR-based cognitive assessments and training [35, 100, 122]. If users find the system disorienting, uncomfortable, or overly complex, it can diminish their willingness to participate and decrease the technology's overall acceptability [48, 93].

Different user groups have unique needs and challenges when engaging with XR technologies. For children, XR provides an engaging and interactive learning environment but raises concerns regarding safety, content appropriateness, and the effects of prolonged exposure to immersive experiences [130]. Middle-aged adults often benefit from XR-based cognitive training and rehabilitation but may face challenges integrating time-intensive XR programs into their daily responsibilities, such as work and family obligations [29]. Flexible and time-efficient XR applications can help address these barriers, improving adherence and overall acceptability [29].

Older adults present another distinct challenge [41]. While XR has shown significant promise for cognitive enhancement in this population, factors such as limited technological familiarity, physical discomfort caused by headsets, and cognitive overload may reduce usability and engagement [22, 48]. Additional support, such as training sessions and user-friendly designs, is often necessary to improve acceptance and ensure meaningful participation [48]. Similarly, individuals with cognitive impairments, including those with dementia or acquired brain injuries, stand to benefit from XR-based cognitive interventions [31, 74]. However, these individuals are often more susceptible to disorientation, cognitive strain, and accessibility challenges, underscoring the importance of carefully designed, adaptive XR systems that cater to their specific needs [11, 41]. Ensuring that XR systems are designed with the specific needs of these populations in mind is critical for promoting user acceptability [2, 35].

### 3.3. Case Studies on UX in Cognitive Training

Case studies have demonstrated the potential of XR technologies to enhance cognitive training outcomes across various populations [32, 130]. In studies involving older adults, immersive VR-based cognitive training programs have shown improvements in memory, attention, and executive function [74, 104]. Participants frequently report higher levels of engagement and enjoyment compared to traditional methods, although challenges related to physical discomfort and usability persist [34, 82].

In the case of neurodevelopmental disorders, the use of transdiagnostic approaches based on functional outcomes to identify the specific needs of each individual is crucial, regardless of diagnostic labels and discrete categories [3]. Case studies implementing this approach focus on personalized and adaptive treatment adjust to the specific needs of each individual [6, 49]. For example, studies involving children with Autism Spectrum Disorder (ASD), XR environments have been used to simulate real-world scenarios, such as visiting a store or attending a movie, where participants can practice social and cognitive skills in a safe, controlled setting [49, 65]. Similarly, XR-based cognitive training has demonstrated promising results for children with Attention Deficit Hyperactivity Disorder (ADHD), improving both cognitive and social functioning through repeated immersive practice [105, 130].

In clinical settings, XR has been employed to develop personalized cognitive training programs for individuals with neurodegenerative conditions or traumatic brain injuries [31, 82]. Participants often report that the immersive nature of XR helps them remain focused and motivated, leading to notable cognitive improvements [5]. However, these studies also highlight the need for simplified systems and additional support to address accessibility and usability concerns [32, 79].

### 4. Multimodal Systems in XR Cognitive Applications

#### 4.1. Overview of Multimodalities: GSR, EEG, ET, Hand Tracking, Body Tracking

Multimodal systems in XR environments integrate various sensory and physiological inputs—such as GSR, EEG, ET, hand tracking, and body tracking—to create a comprehensive understanding of users' cognitive, emotional, and motor states [39]. These systems enable XR applications to capture real-time data across multiple dimensions of human behavior, enhancing the precision, adaptability, and effectiveness of cognitive assessment and training [128].

By leveraging multimodal inputs, XR systems provide deeper insights into user engagement, stress, and cognitive load, facilitating personalized and dynamic experiences [106, 114]. For instance, EEG-based systems can detect mental fatigue and adjust task difficulty accordingly, while GSR data reveals emotional responses that influence performance [115, 125]. Similarly, ET and hand tracking allow for intuitive interactions with virtual environments, improving immersion and usability [17, 110]. A detailed breakdown of key modalities, their descriptions, and applications in XR is presented in Table 1 below.

**Table 1.** Multimodal Systems in XR Cognitive Applications

| Modality | Description | Key Applications in XR |
|---|---|---|
| GSR (Galvanic Skin Response) | Measures the skin's electrical conductivity, which changes with levels of physiological arousal. It is a direct indicator of emotional states such as stress, excitement, or calmness. | Used to track and analyze emotional responses during immersive experiences, such as stress levels during virtual simulations or training exercises. |
| EEG (Electroencephalography) | Records the brain's electrical activity using non-invasive sensors placed on the scalp. It provides real-time data on neural processes related to attention, cognitive workload, and emotional regulation. | Applied in monitoring cognitive load, attention, and engagement levels, especially during tasks requiring high mental effort, such as virtual learning environments or problem-solving scenarios. |
| ET (Eye Tracking) | Monitors and records eye movements, including where and how long a person focuses on specific elements. It helps understand visual attention and perception in XR environments. | Used for evaluating user attention, navigation patterns, and visual processing. Commonly implemented in user interface testing, training simulations, and studies on how users interact with complex visual scenes. |
| Hand Tracking | Detects and interprets hand movements and gestures, allowing for natural and intuitive interaction with virtual objects without the need for handheld controllers. | Enables realistic manipulation of virtual objects, essential for training simulations, virtual prototyping, and enhancing user immersion through gesture-based controls. |
| Body Tracking | Captures full-body movements and postures, providing comprehensive data on physical behavior and motor coordination. It is crucial for assessing how users move and interact within the virtual space. | Utilized in applications that require accurate assessment of motor skills, spatial awareness, or physical training. It's particularly valuable in rehabilitation, sports training, and VR experiences that simulate physical activities. |

#### 4.2. Integration of These Modalities in XR Environments

The integration of multimodal systems into XR environments revolutionizes cognitive assessments and training by enabling continuous, real-time data collection across multiple dimensions of behavior [39]. Unlike traditional methods, which gather isolated data points, XR systems equipped with EEG, GSR, ET, and motion tracking provide a synchronized, dynamic understanding of cognitive and emotional processes as they unfold during task performance [14, 71].

For instance, during a cognitive task, EEG monitors cognitive load, GSR tracks emotional arousal, and ET analyzes attentional focus, while hand and body tracking capture motor performance [77, 110]. This integrated approach offers a holistic view of participants' states, particularly when multiple cognitive functions, such as decision-making and problem-solving, must operate simultaneously [16].

A key innovation enabled by multimodal integration is adaptive feedback. XR environments can dynamically respond to user states by adjusting task parameters in real time [7]. For example, if EEG data indicates cognitive overload or GSR detects elevated stress levels, the system can reduce task complexity to optimize performance and maintain engagement [19, 114]. These adaptive capabilities make XR environments highly responsive and personalized, which enhances their utility in both research and clinical settings [7].

*4.3. Applications of Multimodal Systems in Cognitive Assessment*

Multimodal systems in XR environments enable novel applications for cognitive assessment by combining data from various physiological and behavioral modalities [39]. For instance, emotion recognition benefits from the integration of GSR, EEG, and facial expression analysis, allowing researchers to identify emotional states like stress, frustration, or excitement [106]. Such insights are particularly valuable for mental health assessments, where understanding emotional responses informs personalized therapeutic interventions [101].

Attention monitoring represents another key application. By integrating EEG and ET, XR systems can evaluate attentional control and detect lapses in focus more accurately than traditional assessments [7]. For example, in virtual driving simulations, ET monitors visual focus on key environmental cues, while EEG measures the cognitive effort required to sustain attention [70, 125].

Cognitive load assessment is significantly enhanced through multimodal integration. EEG data, combined with behavioral metrics from hand and body tracking, reveals how individuals manage cognitive demands during complex tasks [81]. This is particularly beneficial in educational and training contexts, where real-time adjustments to task difficulty help optimize learning outcomes and prevent cognitive overload [102].

Finally, multimodal systems are valuable for assessing visuomotor coordination and spatial cognition [113]. By tracking hand and body movements during virtual tasks, XR can evaluate motor-cognitive integration in realistic settings, such as virtual rehabilitation exercises or training simulations [115]. These applications demonstrate how multimodal systems expand the capabilities of XR technologies, enabling assessments that are both more immersive and ecologically valid [123].

*4.4. Advantages and Limitations of Multimodal Systems*

The primary advantage of multimodal systems in XR environments lies in their ability to provide a rich, multidimensional view of cognitive, emotional, and motor processes [39]. By integrating data from EEG, GSR, ET, and motion sensors, these systems offer more accurate assessments and enable personalized interventions that adapt dynamically to users' needs [107]. This holistic approach is particularly beneficial in cognitive training and rehabilitation, where understanding the interplay between cognitive load, emotional states, and motor responses is critical for achieving meaningful outcomes [127].

However, the adoption of multimodal systems is limited by technical complexity and resource demands [9, 100]. Integrating and synchronizing data streams requires sophisticated hardware, software, and computational techniques, which can be costly and inaccessible for resource-limited institutions [109]. Moreover, the sheer volume of data generated can lead to data overload, necessitating advanced analytical methods and expertise for effective interpretation [9].

Physical discomfort caused by wearable sensors and prolonged headset use remains another challenge, particularly for populations with physical or cognitive impairments [78]. Ensuring that systems are lightweight, ergonomic, and inclusive is essential for enhancing UX and expanding the applicability of XR-based multimodal systems [85]. Despite these limitations, the integration of multimodal systems represents a major step forward in cognitive science, offering unparalleled opportunities for real-time, ecologically valid assessments and interventions [77].

**5.    XR Applications in Cognitive Assessment**

*5.1.    Review of Current XR-based Cognitive Assessment Tools*

XR technologies have introduced a new dimension to cognitive assessment by combining immersive realism and dynamic interactivity, allowing researchers to evaluate multiple cognitive domains within real-world-like environments [100]. These tools assess critical cognitive functions, including memory, executive functions, attention, and spatial cognition, through tailored tasks that replicate the complexity of everyday scenarios [10, 83] (see Table 2).

For memory assessment, XR-based tools immerse participants in tasks that require them to remember objects, locations, or sequences within virtual spaces [83]. For instance, the VR Everyday Assessment Lab (VR-EAL) evaluates episodic and prospective memory through tasks like recalling items from a shopping list or remembering to complete actions triggered by time cues [63]. By embedding these tasks in realistic contexts, VR-EAL captures naturalistic memory use, far exceeding the ecological validity of traditional list-based recall tests [63].

In assessing executive functions, XR tools simulate complex multi-step activities that mirror real-world demands [10]. In one example, participants manage tasks within a Virtual Office Simulation by prioritizing assignments, responding to interruptions, and making strategic decisions [51]. Another study used a virtual classroom to evaluate inhibitory control in adults with ASD, demonstrating sensitivity to everyday challenges not captured by conventional tasks [94]. These XR tools provide insights into cognitive flexibility, planning, and adaptability under realistic, high-pressure conditions [10].

Attention assessment in XR environments introduces a higher level of realism by placing participants in immersive, distracting settings [63]. For instance, participants might focus on instructions in a virtual classroom while filtering out noise from peers or other environmental distractions [23, 50]. Such tasks capture attentional control in ways that static, low-stimulation environments cannot [100].

For spatial cognition, XR tasks require participants to navigate virtual cities, plan routes, and interact with 3D objects to evaluate spatial memory and visuospatial reasoning [37, 118]. These tools are particularly useful for assessing populations with spatial deficits, such as individuals with neurodegenerative conditions or brain injuries, offering dynamic, real-world-like scenarios that traditional paper-and-pencil tests fail to replicate [43].

By offering highly realistic, adaptive, and interactive tasks, XR tools provide a deeper, ecologically valid understanding of cognitive abilities across multiple domains, bridging the gap between laboratory settings and real-world functioning [100] (see Table 2).

**Table 2.** XR Cognitive Assessment Tools and Studies

| Cognitive Domain | XR-based Assessment Tool and Study | Description of Method | Key Findings & Implications |
|---|---|---|---|
| Memory | VR-EAL (Kourtesis et al., 2021) [63:20] | Participants engage in tasks like remembering a shopping list or recalling sequences in a realistic virtual environment. | Enhanced ecological validity compared to traditional tests, accurately reflecting real-world memory usage. |
| | Spatial Recall Task (Sauzéon et al., 2016) [108] | Participants memorize and recall spatial information in a virtual environment. | Increased realism leads to better memory performance measurements, compared to static tests. |
| | Context-rich Memory Tasks (Pflueger et al., 2023) [99] | Memory tasks incorporate environmental and situational cues in VR settings. | Contextual elements enhance memory assessment and provide a more realistic understanding of memory function. |
| | VR-EAL (Kourtesis & MacPherson, 2023) [58] | s everyday situations that demand prospective memory, where users must remember tasks triggered by specific times or events, like taking virtual medication after breakfast or at scheduled intervals. | XR methods outperform traditional approaches in capturing prospective memory in real-life situations. |
| Executive Functions | Virtual Office Simulation (Jansari et al., 2014) [51] | Participants manage tasks, handle unexpected events, and make strategic decisions in a virtual office setting. | Effectively assesses planning, adaptability, and decision-making, mirroring real-world complexities. |
| | Inhibitory Control in ASD (Parsons & Carlew, 2016)[94] | VR classroom simulation to measure inhibitory control in adults with ASD. | Captures real-world executive dysfunction in a way that traditional tests cannot. |
| | VR-EAL (Kourtesis & MacPherson, 2021) [63:20] | Tasks simulate planning and adaptability challenges, like running errands in a virtual city. Also, there is a cooking task which requires multi-tasking skills. | Provides insights into strategic planning and adaptability under realistic conditions. |
| Attention | High-Stimulation Attention Task (Coleman et al., 2019) [23] | Participants focus on instructions amid distractions in a virtual classroom. | More accurate assessment of attention control compared to lab-based tests. |
| | Naturalistic Attention (Iriarte et al., 2016 [50] | Participants filter out distractions in an immersive VR class. | XR tasks simulate real-world attentional demands, offering more applicable results. |
| | VR-EAL (Kourtesis et al., 2021) [63:20] | Detecting visual/auditory cues amid distractions while on the road. | Comprehensive assessment of attentional processes, enhancing real-world applicability. |

| Cognitive Domain | XR-based Assessment Tool and Study | Description of Method | Key Findings & Implications |
| --- | --- | --- | --- |
| Visuospatial Skills | Virtual City Navigation (Grübel et al., 2017) [37] | Participants plan routes and remember landmarks in a virtual city. | Detailed data on spatial memory and reasoning that traditional 2D tests cannot offer. |
| | Spatial Deficit Assessment (Howett et al., 2019) [43] | XR tasks assess navigation skills in individuals with mild cognitive impairment or brain injuries. | Valuable for clinical applications, as XR provides a realistic measure of spatial impairments. |
| | VR-EAL (Kourtesis & MacPherson, 2021) [63:20] | Route planning and landmark recall in immersive VR settings. | Offers an ecologically valid measure of spatial reasoning, closely reflecting real-world challenges. |
| | 3D Interaction Tasks (Cogné et al., 2018) [21] | Participants interact with 3D objects in virtual environments to test coordination and movement patterns. | XR captures coordination skills in a dynamic setting, revealing nuances not measurable by traditional tests. |
| | Object Manipulation (Wen et al., 2023b) [127] | Tasks involving manipulation of objects and solving spatial puzzles. | Provides a comprehensive understanding of visuomotor skills in realistic, engaging scenarios. |

### 5.2. XR Assessments Compared to Traditional Methods

The key distinction between XR-based cognitive assessments and traditional methods lies in XR's ability to simulate real-world complexity and provide an engaging, immersive testing environment [100]. Traditional cognitive assessments, such as computerized tasks, structured interviews, or paper-and-pencil tests, typically isolate specific cognitive functions—such as memory, attention, or problem-solving—in controlled, low-stimulation settings [44]. While these standardized methods provide valuable, quantifiable data, they fail to replicate the dynamic and contextual demands of real-world cognition, where tasks are rarely performed in isolation or under ideal conditions [63, 100].

#### 5.2.1. Ecological Validity and Realism

XR-based assessments excel in ecological validity by immersing participants in realistic, interactive scenarios that simulate everyday tasks [95]. For example, while traditional memory tasks might involve recalling a list of words or numbers, XR environments require participants to remember object locations or execute sequential instructions within immersive, dynamic spaces [99, 108]. Tasks such as navigating through a virtual house, remembering the placement of items, or recalling event-based triggers provide a naturalistic evaluation of memory processes, making the findings more transferable to real-world settings [83, 127].

#### 5.2.2. Visuomotor Coordination and Spatial Cognition

Another significant advantage of XR assessments is their ability to measure visuomotor coordination and spatial cognition more effectively than traditional methods [113]. Standard tests might involve drawing figures or solving two-dimensional mazes [28]. In contrast, XR tasks allow participants to interact with 3D objects, solve spatial puzzles, and navigate immersive environments that replicate real-world spatial challenges [21]. This dynamic approach captures nuances of visuospatial reasoning and coordination that are difficult to assess using static, two-dimensional tools [127]. Additionally, XR tasks often provide real-time feedback, enabling researchers to observe how participants adjust their strategies in response to task demands [101].

5.2.3. Continuous and Dynamic Data Collection

A critical advantage of XR-based assessments is their capacity for continuous data collection, which offers a more detailed and dynamic understanding of cognitive performance [7, 127]. Traditional assessments typically measure outcomes at discrete time points—such as reaction times, accuracy, or task completion scores—providing limited insight into the processes underlying performance [44]. In contrast, XR systems collect continuous streams of behavioral, physiological, and neural data throughout the task [115].

For instance, an XR-based attention task can simultaneously track participants' gaze (using ET), their emotional responses (via GSR), and their cognitive load (via EEG) [114]. This approach reveals not only whether participants complete the task successfully but also how their focus shifts, how cognitive demands fluctuate, and how emotional states impact performance [77]. Such continuous, multimodal data enable researchers to derive nuanced insights into the dynamic interplay of cognitive and emotional processes [7].

5.2.4. Complementarity with Traditional Methods

Despite their advantages, XR-based assessments do not aim to replace traditional cognitive testing methods but rather to complement them [100]. Traditional tools remain valuable for standardization and benchmarking, particularly in clinical and research settings where consistency and simplicity are critical [44]. The integration of XR with traditional assessments can provide a comprehensive evaluation of cognitive abilities, combining the strengths of standardized methods with the ecological validity and richness of XR technologies [15]. By offering ecologically valid, dynamic, and multimodal assessments, XR tools address key limitations of traditional methods, providing a deeper and more contextually relevant understanding of cognitive functioning in everyday life [63].

5.3. *XR's Potential for Real-time Data Collection and Analysis*

XR-based cognitive assessments offer a significant advantage through real-time, multimodal data collection [101]. Unlike traditional assessments, which rely on isolated performance metrics, XR systems continuously collect synchronized data streams, including:
- Behavioral data: Eye movements, reaction times, and body posture [77, 110].
- Physiological data: Stress or arousal measured via GSR [19].
- Neural data: Cognitive states assessed through EEG signals [126].

This multimodal approach enables a deeper understanding of cognitive and emotional processes as they unfold during task performance [106]. For instance, an XR-based decision-making task can simultaneously track attention (ET), emotional responses (GSR), and cognitive load (EEG), revealing how these processes interact under complex conditions [36, 77].

Additionally, XR systems enable adaptive assessments by dynamically adjusting task difficulty based on real-time feedback [129]. For example, if EEG or GSR data detects cognitive overload, the system can lower task complexity to prevent frustration [19]. Conversely, if the task is too easy, complexity can increase to maintain engagement [129]. This responsive, personalized feedback ensures that XR tools remain both engaging and effective [9].

5.4. *Challenges in Implementing XR in Large-scale Cognitive Assessments*

Despite their transformative potential, XR-based cognitive assessments face several challenges that hinder their widespread implementation, particularly at scale [100]. One of the most significant barriers is the cost and resource requirements associated with XR systems [100]. High-quality hardware, such as VR and AR headsets, sensors, and the powerful computers needed to run immersive environments, remains relatively expensive compared to traditional cognitive assessment tools [101]. Moreover, developing scientifically valid and engaging XR-based applications requires specialized expertise in cognitive science, software development, and XR design, which can be both costly and resource-intensive [109]. For institutions with limited financial and technical resources, these requirements can present substantial obstacles to adoption and maintenance [61].

Another major challenge is usability and accessibility, particularly for specific populations, including older adults, individuals with physical impairments, or those with cognitive disabilities [69, 112]. While XR systems provide unparalleled immersion, prolonged use can lead to cybersickness, eye strain, and physical discomfort, which can negatively impact user engagement and data reliability [62, 66, 87]. These limitations are particularly concerning for long-duration tasks or assessments targeting vulnerable populations [62, 92]. Designing XR tools that minimize these side effects and are accessible to users with diverse needs is essential for broader adoption [41, 92, 93].

Scalability also poses a significant hurdle for XR technologies, particularly when transitioning from controlled research settings to larger, real-world applications [100]. Administering XR-based cognitive assessments to large populations requires not only an adequate number of XR systems but also the technical infrastructure and support necessary to manage and troubleshoot these devices during testing sessions [109]. Furthermore, XR assessments generate large and complex data sets that require advanced analytical tools and expertise to process effectively [24]. Institutions with limited infrastructure may struggle to manage these demands, further complicating efforts to scale XR tools for widespread use [61, 109].

Nevertheless, advancements in XR hardware and software are expected to alleviate some of these challenges [101]. As the cost of XR devices continues to decline, and technologies become lighter, more ergonomic, and user-friendly, broader accessibility will likely follow [100]. Additionally, emerging cloud-based and AI solutions for XR applications can help reduce on-site computational requirements, making these systems more feasible for resource-limited settings [9]. Addressing these barriers—through affordable hardware, inclusive design, and scalable infrastructure—will be crucial for realizing XR's full potential as an innovative tool for large-scale cognitive assessment [96, 100].

## 6. XR Applications in Cognitive Training

### 6.1. Review of Current XR-based Cognitive Training Interventions

XR technologies have emerged as powerful tools for cognitive training, offering immersive and interactive environments that enhance engagement, personalization, and ecological validity [101]. Unlike traditional training methods, which often rely on repetitive and static tasks, XR enables the creation of dynamic, real-world scenarios tailored to train specific cognitive functions such as memory, attention, and executive function [32].

Memory enhancement is a major focus in XR-based cognitive training [53]. For example, XR tasks immerse participants in context-rich scenarios that simulate real-world challenges, such as remembering object locations within a virtual home or recalling action sequences in a virtual kitchen [53, 120]. These tasks engage memory processes in a way that reflects daily-life challenges, making them more applicable than traditional route learning or list-recall exercises [40]. Studies have shown that such immersive tasks lead to significant improvements in memory performance, particularly among older adults and individuals with mild cognitive impairments (MCI) [88].

Attention training has also benefitted from XR's ability to simulate complex, high-stimulation environments [124]. XR interventions challenge users to sustain attention, filter distractions, and respond to multiple sources of stimuli—skills critical for real-world functioning [46]. For instance, tasks set in virtual marketplaces or classrooms require participants to focus on key details while ignoring irrelevant inputs, providing a more nuanced training experience than simple reaction-time exercises [79, 124]. Additionally, XR systems can incorporate adaptive difficulty and real-time feedback, ensuring that users are continuously engaged at an optimal level [109].

XR has also proven effective in training executive functions, such as problem-solving, planning, and task-switching [74]. Participants may engage in tasks that require navigating a virtual city, managing a simulated workplace, or solving dynamic problems in changing scenarios [74, 89]. These immersive tasks challenge participants to prioritize actions, adapt to unexpected events, and execute strategic decisions—skills that are often difficult to isolate and train in static environments [42]. By providing engaging, interactive, and ecologically valid scenarios, XR-based cognitive training offers a more holistic approach to improving cognitive functions across diverse populations [32].

**Table 3.** XR Cognitive Training Tools and Studies

| Training Focus | Population (Study) | Method Description | Key Findings and Implications |
|---|---|---|---|
| Memory Training | Older Adults (Varela-Aldás et al., 2022) [120] | Real-life simulated memory tasks, like recalling sequences of actions in a virtual kitchen. | Enhanced user engagement and better real-world applicability compared to static recall exercises. |
| | Individuals with Cognitive Decline (Mondellini et al., 2018) [88] | Context-rich scenarios replicating everyday memory challenges. | Memory performance showed marked improvements, especially in older adults and those with MCI. |
| Attention Training | General Population & Stroke Patients (Huygelier et al., 2022) [46] | Dynamic tasks in XR requiring sustained attention in realistic, immersive settings. | Improved attentional control, better reflecting real-world demands compared to simple reaction-time exercises. |
| | General Population, Children (Wang et al., 2020) [124] | Tasks designed with adaptive difficulty and real-time feedback to sustain attention. | Participants maintained engagement and showed greater attentional improvements that generalized to daily activities. |
| | Older Adults (Lorentz et al., 2023)[79] | Immersive attention training tasks set in complex environments, like virtual markets. | Enhanced focus and attentional resource management in high-stimulation scenarios. |
| | General Population & Adults with ADHD (Selaskowski et al., 2023) [109] | XR-based interventions with personalized difficulty adjustments. | Greater effectiveness in training attention skills compared to non-adaptive methods. |
| Executive Functions Training | General Population and MCI Liao et al., 2019) [74] | Participants navigate a virtual city or manage tasks in a simulated workplace, engaging executive functions. | XR tasks provided a more realistic training experience, leading to better problem-solving and adaptability. |
| Social Cognition Training | Adults with ASD (Kourtesis et al., 2023) [65] | Simulations of daily life tasks, like job interviews and shopping, for real-world social skill practice. | XR provided a safe space to learn and adapt, enhancing social interactions and everyday functioning. |
| | Children with ASD (Bekele et al., 2016) [6] | XR scenarios focusing on social interactions, like making eye contact and understanding social cues. | Effective at reducing social anxiety and improving communication skills in a safe, controlled environment. |
| | Children with ASD (Ip et al., 2018) [49] | Virtual practice of social tasks, tailored to individual needs, with repeated exposure. | Personalized training showed significant improvements in social cognition and adaptive behavior. |

| Training Focus | Population (Study) | Method Description | Key Findings and Implications |
|---|---|---|---|
| Multiple Cognitive Domains Training | TBI Patients (Masoumzadeh & Moussavi, 2020) [84] | Gradually increasing task complexity in XR settings to support cognitive skill recovery. | Effective in enhancing spatial memory and task-switching, critical for neurological recovery. |
| | Children with Attention or Learning Challenges (Coleman et al., 2019) [23] | Game-like XR scenarios for working memory, problem-solving, and attention training. | High engagement and sustained interest, resulting in cognitive gains and improved academic skills. |
| | Children (Araiza-Alba et al., 2021) [1] | Interactive missions and virtual puzzles that require strategic thinking and memory use. | Enhanced cognitive skill development and positive behavioral outcomes in young learners. |
| | Children with ADHD (Ou et al., 2020) [91] | XR-based training that focuses on attention and strategic thinking through playful scenarios. | XR tasks promoted adaptability, patience, and academic success. |
| | Children with ADHD (Wong et al., 2023) [130] | Engaging XR tasks for attention, social cognition, and executive function, with adaptable challenges. | Increased focus, better task management, and improved social skills, proving XR to be a highly effective therapeutic tool. |

*6.2. Population-specific XR Training Programs*

One of XR's greatest strengths lies in its ability to adapt to the unique needs of specific populations [123]. This flexibility enables the design of tailored interventions that address the cognitive challenges faced by children, older adults, and individuals with neurodevelopmental or neurological conditions [49, 74].

XR-based cognitive training focuses on enhancing attention, working memory, and problem-solving skills through game-like, interactive scenarios [1, 23]. Tasks such as navigating virtual mazes, solving puzzles, or completing missions not only improve cognitive performance but also foster persistence, patience, and adaptability—skills essential for academic success [91]. The engaging and playful nature of XR makes it particularly effective for sustaining children's motivation and participation [103].

For aging adults, XR-based training addresses cognitive decline by targeting memory, attention, and executive functions through tasks that replicate daily-life scenarios [12]. Examples include remembering shopping lists in virtual stores, preparing meals, or navigating public transport systems [75]. These practical tasks not only improve cognitive performance but also enhance confidence and quality of life in older adults [117]. XR environments can also accommodate physical limitations, ensuring safety and comfort for participants with reduced mobility [98].

For individuals with brain injuries or neurodevelopmental disorders, XR provides a safe and controlled space to practice cognitive skills critical for recovery and daily functioning [123]. In traumatic brain injury (TBI) rehabilitation, XR environments can simulate real-world scenarios—like driving or workplace tasks—that challenge cognitive abilities while allowing gradual increases in task complexity [84]. Similarly, for individuals with ASD, XR-based interventions focus on improving social cognition, executive function, and adaptive skills [6, 49]. Virtual environments simulate real-world social interactions, such as job interviews or shopping, offering a safe space for practice without fear of judgment or consequences [65].

XR's adaptability also benefits children with ADHD. Engaging XR tasks that train attention, social cognition, and executive function have been shown to enhance focus, task management, and behavioral outcomes [130]. The structured yet flexible nature of XR makes it an effective, user-centered therapeutic tool that can be tailored to individual progress and needs [130]. Overall, XR technologies allow for the development of personalized and contextually relevant interventions that address the unique challenges faced by diverse populations, improving cognitive and adaptive skills in ways that are highly transferable to real-world settings [96].

### 6.3. Long-term Effects and Retention of Cognitive Skills in XR Training

The long-term effects and retention of cognitive improvements following XR-based cognitive training are key areas of ongoing research [8]. XR technologies have demonstrated strong potential to produce lasting gains by simulating real-world challenges, enhancing engagement, and fostering deeper learning processes [42, 120]. However, the degree to which these skills transfer to daily tasks and remain stable over time without reinforcement remains a central question [123].

A primary contributor to retention is XR's ecological validity—its ability to replicate realistic tasks [12]. Unlike traditional training, XR tasks simulate daily activities, such as navigating environments, recalling object locations, or solving multi-step problems [26]. These context-rich experiences make cognitive improvements more likely to generalize to real-world situations, as shown in studies where older adults sustained memory gains weeks after completing XR-based training [88, 123].

User engagement also plays a pivotal role in reinforcing long-term cognitive benefits [122]. XR environments provide immersive, interactive experiences that sustain motivation and emotional arousal, key factors for durable learning [117, 131]. Features such as adaptive task difficulty and real-time feedback ensure participants remain appropriately challenged, enhancing both skill acquisition and retention [116].

Nevertheless, challenges remain. While XR training shows promise, some skills may not fully transfer to non-virtual settings, particularly without follow-up reinforcement [12]. Additionally, the optimal duration, frequency, and reinforcement strategies for maintaining long-term gains are still under investigation [123]. Research into XR's role in neuroplasticity—the brain's ability to reorganize and adapt—suggests that engaging in adaptive, cognitively demanding XR tasks can strengthen neural connections, particularly in aging populations or individuals recovering from brain injuries [52]. Future studies should explore these mechanisms to better understand the long-term biological and behavioral impacts of XR-based training [123].

### 6.4. Future Directions in XR-based Cognitive Training

The future of XR-based cognitive training will be shaped by advancements in technology, personalization, and accessibility, offering new possibilities for enhancing cognitive health across diverse populations [123].

One promising direction is the integration of XR with neuroimaging techniques such as EEG [127]. Real-time monitoring of brain activity can provide insights into users' cognitive states, allowing systems to dynamically adapt task difficulty based on cognitive load or fatigue [9, 119]. This combination of XR and neuroimaging represents a significant step toward creating tailored interventions that optimize learning outcomes [126].

Advances in socially interactive XR environments will further expand the scope of cognitive training [65]. Virtual environments featuring avatars and collaborative scenarios can simulate real-world challenges that require users to develop problem-solving, emotional regulation, and social cognition skills [38]. These interactive platforms are particularly beneficial for individuals with ASD or social anxiety, offering safe, repeatable spaces to practice and refine social behaviors [6, 65].

Artificial Intelligence (AI) will also revolutionize XR training. AI-driven algorithms can analyze user performance data to predict cognitive states, personalize training trajectories, and adjust difficulty in real time, ensuring participants remain challenged and engaged [9]. AI can further enhance XR environments by enabling dynamic, naturalistic responses to user actions, creating immersive experiences that closely replicate real-world cognitive demands [119].

Finally, as XR technologies become more affordable and accessible, their adoption will expand into resource-limited settings such as schools, clinics, and homes [101]. The development of lightweight, ergonomic hardware and cloud-based platforms will enable remote delivery of XR training programs, democratizing access to high-quality cognitive interventions for underserved populations [11, 101].

These advancements will transform XR-based cognitive training into a scalable, personalized, and effective solution for addressing cognitive challenges. Future research should prioritize validating these approaches through longitudinal studies and exploring their potential for lifelong cognitive health, rehabilitation, and development [8, 122].

## 7. Clinical Utility of XR in Cognitive Assessment and Training

### 7.1. Benefits of XR-based Cognitive Tools in Clinical Settings

XR technologies—VR, AR and MR—offer unique advantages in clinical settings for cognitive assessment and training [101]. XR provides immersive, interactive, and flexible platforms that enable the evaluation and improvement of cognitive functions in ways that traditional methods cannot [32, 101]. By simulating real-world scenarios, XR enhances ecological validity, bridging the gap between laboratory-based testing and real-life performance [100, 128].

In clinical practice, XR tools allow clinicians to observe cognitive processes, such as memory, attention, and executive functions, in dynamic, controlled environments [10, 53]. For instance, memory assessments can involve navigating virtual spaces and recalling item locations, providing richer insights compared to static tasks [47, 99]. Similarly, XR-based tasks can simulate multitasking challenges, enabling clinicians to evaluate decision-making and problem-solving under realistic conditions [10, 63]. This comprehensive approach is particularly valuable for individuals with cognitive impairments, such as dementia or TBI, as it reveals how they manage complex tasks, distractions, and spatial challenges in real time [111].

Another major advantage of XR tools is their customizability and adaptability [131]. Task difficulty can be adjusted in real time, allowing clinicians to personalize assessments and training to meet individual needs [7, 131]. This adaptability enhances patient engagement and adherence, which are critical for achieving meaningful improvements, particularly among older adults and children [93, 103].

### 7.2. Comparative Analysis of Traditional vs. XR-based Cognitive Assessments and Training

Traditional cognitive assessments—such as paper-and-pencil tests or computerized tasks—are valuable for standardized evaluation but often lack ecological validity and fail to reflect real-world cognitive demands [44, 95]. XR-based assessments, in contrast, immerse participants in realistic, interactive environments that engage multiple cognitive processes simultaneously [100].

For example, traditional memory tasks may involve recalling word lists, whereas XR tools place participants in virtual environments—like grocery stores—where they must remember and retrieve items, integrating attention, memory, and spatial navigation [63, 95]. Similarly, XR enables real-time collection of multimodal data—ET, motion tracking, and physiological signals—providing deeper insights into participants' cognitive and emotional states during task performance [18].

In cognitive training, XR tools outperform traditional repetitive exercises by offering engaging, context-rich tasks [123]. For instance, XR-based interventions may simulate driving, workplace activities, or social interactions, enhancing motivation and facilitating skill transfer to real-world situations [49, 84]. Unlike static methods, XR dynamically adapts difficulty based on user performance, ensuring continuous engagement and personalized progression [89].

### 7.3. Case Examples of Clinical Applications

XR has demonstrated significant clinical utility across various populations and conditions [20, 32]. For individuals with MCI or dementia, XR tools simulate daily activities—such as navigating virtual cities or managing household tasks—to assess spatial memory, attention, and executive function [43, 76]. These immersive assessments provide accurate, ecologically valid insights that aid in early diagnosis and personalized care plans [76].

In rehabilitation settings, XR has been employed to improve cognitive and motor recovery for stroke or TBI patients [111]. Tasks such as reaching for virtual objects, navigating spaces, or multitasking allow clinicians to gradually increase task complexity while tracking progress in real time [74, 80].

For individuals with ASD, XR enables safe and structured practice of social cognition and communication skills [49]. Virtual avatars and interactive environments help individuals engage in realistic social tasks—such as making eye contact, participating in conversations, or attending job interviews—without the pressures of real-world interactions [65].

XR tools have also shown promise for older adults, offering cognitive training scenarios like remembering appointments or managing finances within virtual environments [90]. These tasks improve cognitive function, build confidence, and provide a non-invasive, engaging approach to delaying cognitive decline [47, 90].

### 7.4. Barriers to XR Adoption: Hardware, Software, and Accessibility Challenges

While XR technologies hold immense potential for cognitive assessment and training, several barriers continue to hinder their widespread adoption [8]. A major challenge is the cost of XR hardware, including high-quality headsets, sensors, and computing systems, which remain prohibitively expensive for many educational, clinical, and research institutions, particularly in low-resource settings [100, 101]. Although hardware prices have decreased in recent years, the financial burden associated with acquiring and maintaining XR systems remains significant [109].

Developing XR software is another resource-intensive barrier [64]. Creating engaging, user-friendly applications requires substantial expertise, financial investment, and continuous updates to ensure compatibility with rapidly evolving hardware technologies [96, 101]. Institutions lacking the necessary infrastructure or funding may struggle to integrate XR effectively into cognitive programs [109].

Accessibility also poses significant challenges. Operating XR systems requires a certain level of technological literacy, which may not be present in older adults or individuals with limited experience using digital technologies [48]. Physical barriers, such as discomfort caused by prolonged headset use or difficulties interacting with virtual environments due to motor impairments, further complicate adoption [30, 64]. Additionally, individuals with visual or hearing impairments may face challenges if XR systems are not designed with adaptive accessibility features [85, 93].

Additionally, the technical complexity of XR implementation—developing, maintaining, and integrating these systems into existing clinical workflows—requires specialized expertise in cognitive science and software engineering [9]. Clinician training and system compatibility with other diagnostic tools are further obstacles to adoption [109].

Regulatory issues surrounding data privacy and security also pose challenges. XR systems collect sensitive biometric and behavioral data, necessitating clear frameworks to ensure compliance with clinical standards for patient safety and data protection [15, 112].

Overcoming these barriers requires a concerted effort to make XR systems more affordable, user-friendly, and inclusive [100, 101]. Prioritizing streamlined software, lightweight hardware, and adaptive design will be key to ensuring that XR technologies are accessible and acceptable across diverse populations [30].

### 7.5. Current Issues with Using XR for Cognitive Assessment and Training

### 7.5.1. Cybersickness and Immersion Fatigue

One of the primary challenges of XR technologies is cybersickness, a phenomenon similar to motion sickness caused by mismatches between visual input and vestibular signals [8, 56]. Symptoms such as nausea, dizziness, and eye strain can limit how long participants can engage with XR tasks and negatively impact cognitive performance during assessments and training [62, 87]. Cybersickness is particularly problematic in applications requiring prolonged immersion or rapid movements, such as navigation tasks or VR-based simulations [62, 68].

In addition to cybersickness, immersion fatigue can occur when participants are exposed to highly immersive environments for extended periods [62]. While XR's immersive nature enhances engagement, excessive cognitive and sensory stimulation may cause mental fatigue, visual discomfort, and declining task performance over time [24, 114]. This challenge is especially relevant for populations with cognitive impairments or older adults, where sustained focus may already be limited [20]. Optimizing session durations, implementing periodic breaks, and managing the intensity of tasks are essential strategies for mitigating these effects [62, 80].

### 7.5.2. Underutilization of XR Technologies

Despite technological advancements, many XR applications for cognitive assessment and training fail to fully exploit XR's capabilities [101]. A significant area of underutilization lies in the integration of multimodal data streams, such as combining visual, auditory, and haptic feedback with physiological measures like EEG, GSR, and heart rate variability [9, 86]. Multimodal integration allows for more adaptive, personalized experiences and provides deeper insights into users' cognitive and emotional states [77].

For example, incorporating haptic feedback can make interactions with virtual objects more natural [59, 121], while real-time monitoring of physiological responses can dynamically adjust task difficulty to maintain an optimal cognitive load [19]. However, many current XR systems focus primarily on visual and auditory inputs, missing opportunities to enhance immersion, interactivity, and the granularity of cognitive assessments [57]. Expanding the use of these modalities could significantly improve both the UX and the accuracy of XR-based cognitive and training tools [9, 57].

### 7.5.3. Population-Specific Challenges

The effectiveness and usability of XR technologies can vary significantly across different populations, creating unique challenges for their widespread adoption.

- Children: While XR holds great promise for cognitive training in younger populations, children may be more susceptible to cybersickness due to their developing vestibular systems [103]. Additionally, children require highly interactive, engaging content to sustain their attention, making careful design of XR tasks essential [103, 130].
- Older Adults: Older populations often face barriers related to technology adoption. Physical discomfort caused by heavy or poorly balanced headsets, visual fatigue, and unfamiliarity with immersive interfaces can limit their participation and effectiveness [41, 93]. Tailoring XR experiences to accommodate physical and cognitive limitations is necessary to make these tools accessible to aging adults [112].
- Cognitive Impairment: Individuals with dementia, TBI, or other neurological conditions may find XR environments overwhelming or disorienting due to their immersive nature and the cognitive load required to navigate these systems [2, 11]. Designing XR tools with simplified interfaces, adjustable immersion levels, and clear guidance can help address these challenges [20].

Addressing population-specific barriers requires thoughtful, user-centered design that prioritizes accessibility, comfort, and usability across diverse user groups [8, 101].

### 7.5.4. Hardware Limitations and Their Impact on Immersive Experience

The quality of the XR experience is heavily dependent on the underlying hardware, and current limitations remain a significant hurdle for clinical applications [61]. High-quality XR systems rely on high-resolution displays, wide fields of view, and responsive motion tracking, but limitations in these areas can reduce immersion and user comfort [101]. For example, low-resolution visuals can cause pixelation, while latency in motion tracking can disrupt natural interactions with virtual objects, reducing the accuracy and realism of cognitive tasks [81].

The weight and ergonomics of VR headsets are also critical concerns [41]. Heavy or poorly balanced headsets can lead to physical discomfort, particularly during prolonged use, which limits session duration and may reduce user engagement [93]. These challenges are exacerbated for populations such as older adults and individuals with physical impairments [48]. Future advancements in hardware design—such as lighter, more ergonomic devices with improved motion tracking and display performance—are essential for enhancing usability and expanding XR's clinical utility [85, 101].

### 7.5.5. Strategies to Mitigate Challenges and Improve XR Implementation

Several strategies can address these challenges to maximize the effectiveness of XR technologies in cognitive assessment and training [69]. To reduce cybersickness, techniques such as dynamic field-of-view adjustment, optimizing frame rates, and improving motion tracking accuracy can minimize sensory mismatches [33, 62]. For managing immersion fatigue, task designs that incorporate breaks, natural transitions, and varied cognitive loads can help sustain engagement without overwhelming users [80].

Addressing underutilization of XR technologies requires expanding the integration of multimodal data streams and sensory feedback, such as EEG, GSR, and haptics, to provide more adaptive, personalized cognitive tools [9, 19]. Thoughtful design that accounts for population-specific needs—including intuitive interfaces, adjustable immersion levels, and physical accommodations—will enhance accessibility and usability for diverse groups [41, 103]. [85, 101]

Finally, continued advancements in hardware—including lighter, more ergonomic headsets with improved resolution and tracking capabilities—will be critical for delivering immersive, comfortable experiences that support effective cognitive assessment and training [85, 101].

### 7.5.6. Guidelines to Maximize the Effectiveness of XR Tools for Assessment and Training.

XR experts and workgroups are developing guidelines and checklist to ensure that XR assessment and training applications meet optimal criteria and research rigor [67, 69, 122]. XR tools for cognitive assessment employ a multidimensional checklist to ensure successful development, with a focus on ecological relevance, task adaptability, and anticipating predictable pitfalls [67, 69]. Recent guidelines, address the necessity for a multidisciplinary workgroups in the XR development applications and emphasizing on the integration of multimodal techniques, to ensure the VR applications conduct an adequate cognitive assessment [64, 67, 80].

Additionally, growing efforts are focused on developing guidelines to analyze XR rehabilitation applications in the early stages of development, ensuring optimal design, rigorous protocol testing, and comprehensive evaluation of human factors such as acceptability, usability, cybersickness, and safety [122]. For example, a recently proposed framework introduces AI techniques to adapt task difficulty and personalize the VR training process through an adaptive VR application [80]. Constantly updating guidelines and frameworks is crucial to keep pace with continuous technological advancements and software development [69, 122].

## 8. Conclusions

This review underscores the transformative potential of XR technologies in cognitive assessment and training, particularly in their ability to provide ecologically valid evaluations of real-world cognitive skills. Unlike traditional methods, XR immerses participants in dynamic, interactive environments, enabling a comprehensive analysis of memory, attention, decision-making, and problem-solving as they occur in everyday contexts.

A defining strength of XR is its capacity for multimodal integration. Systems combining GSR, EEG, ET, hand tracking, and body tracking allow for a deeper, holistic understanding of users' cognitive and emotional states. These innovations have the potential to drive adaptive and personalized interventions, yet many current XR applications fail to fully exploit this capability. Future advancements in multimodal systems will be critical for enhancing both engagement and therapeutic outcomes.

XR's engaging and interactive environments are particularly advantageous for cognitive training programs, fostering motivation and adherence that are often lacking in traditional approaches. However, challenges such as cybersickness, immersion fatigue, and hardware constraints remain significant barriers to widespread adoption. Addressing these issues through ergonomic hardware design, intuitive interfaces, and enhanced software optimization will be essential for ensuring accessibility and comfort for diverse users.

The validation of XR-based rehabilitation programs through rigorous randomized controlled trials (RCTs) is another pressing requirement. Such trials must be supported by multicenter collaborations, standardized protocols, and iterative testing methodologies to establish both scientific reliability and practical applicability. Early-phase studies and continuous feedback from diverse user groups can further refine these technologies for specific clinical and educational settings.

Despite these advances, XR technologies face substantial limitations. These include the underutilization of multimodal features, usability challenges for older adults and cognitively impaired populations, and regulatory hurdles concerning data privacy and ethical standards. Resolving these issues is essential for achieving the broader acceptance of XR in healthcare and education.

Looking ahead, the future of XR lies in its ability to seamlessly integrate cutting-edge technologies, enhance user-centric designs, and expand its applicability to both general and specialized populations. By overcoming current limitations, XR can solidify its role as a transformative tool in neuropsychology and cognitive science, offering innovative and impactful solutions for assessment, training, and rehabilitation.

*8.1. Future Directions*

To realize the full potential of XR in cognitive science, future advancements must address critical areas to overcome existing limitations and expand its transformative capabilities. One essential focus is the enhancement of multimodal integration. By incorporating diverse biometric data streams—including EEG, GSR, and ET—XR systems can enable real-time task adaptations and provide a more immersive and responsive user experience. This integration is pivotal for improving the accuracy and effectiveness of cognitive training and assessments.

Equally significant is the development of personalized and adaptive solutions tailored to the unique cognitive and emotional needs of individual users. Such customization is particularly valuable for vulnerable or clinical populations, where generic approaches often fall short. By leveraging real-time feedback and data analysis, XR systems can dynamically adjust their content to align with users' specific requirements, enhancing both engagement and outcomes.

Advances in hardware design are also paramount. The development of lighter, more ergonomic headsets and more precise motion tracking systems will not only improve usability but also expand accessibility to broader audiences. These innovations are crucial for mitigating issues such as physical discomfort and usability barriers, making XR technologies more viable for prolonged use in educational, clinical, and domestic settings.

Additionally, XR's ecological validity opens new opportunities for its application in diverse, real-world scenarios. From navigation and workplace simulations to household management, XR tools can provide invaluable insights into users' cognitive abilities in practical contexts. These applications extend XR's utility beyond traditional domains, offering solutions that are both innovative and pragmatic.

Rehabilitation is another area where XR's flexibility and adaptability shine. By creating safe, controlled environments, XR enables individuals to practice real-world tasks critical to their recovery. This capability is particularly beneficial for patients managing neurological injuries or cognitive impairments, as it fosters consistent engagement and improves therapeutic outcomes. The ability to adapt these environments to individual needs ensures that rehabilitation programs are both effective and accessible.

Ultimately, addressing these priorities will drive the evolution of XR technologies into inclusive, robust, and impactful tools. By overcoming current limitations, XR can solidify its role as a cornerstone in neuropsychology and cognitive science, facilitating groundbreaking advancements in research, education, and clinical practice.

**Supplementary Materials:** No Supplementary Materials.

**Author Contributions:** Conceptualization, P.V.G.E. and P.K.; validation, P.V.G.E., S.F.G. and P.K..; writing—original draft preparation, P.V.G.E. and P.K..; writing—review and editing, P.V.G.E., S.F.G. and P.K.; supervision, S.F.G. and P.K.; project administration, S.F.G. and P.K.. All authors have read and agreed to the published version of the manuscript.

**Funding:** This research received no external funding.

**Institutional Review Board Statement:** Not applicable.

**Informed Consent Statement:** Not applicable.

**Data Availability Statement:** Data sharing is not applicable.

**Conflicts of Interest:** The authors declare no conflicts of interest.

Cognitive Assessment and Training in Extended Reality                                                                                                         21